\documentclass[preprint,showpacs,preprintnumbers,amsmath,amssymb]{revtex4}
\usepackage{graphicx}
\usepackage{dcolumn}
\usepackage{bm}

\begin{document}

\title{Instantonic Methods for Quantum Tunneling in Finite Size}
\author{ Marco Zoli }
\affiliation{
School of Science and Technology - CNISM, University of Camerino, 62032 Camerino, Italy. - marco.zoli@unicam.it}

\date{\today}

\begin{abstract}
The instantonic approach for a $\phi^4$ model potential is
reexamined in the asymptotic limit. The path integral of the system
is derived in semiclassical approximation expanding the action
around the classical background. It is shown that the singularity
in the path integral, arising from the zero mode in the quantum
fluctuation spectrum, can be tackled only assuming a {\it finite}
(although large) system size. On the other hand the standard
instantonic method assumes the (anti)kink as classical background.
But the (anti)kink is the solution of the Euler-Lagrange equation
for the {\it infinite} size system. This formal contradiction can
be consistently overcome by the finite size instantonic theory
based on the Jacobi elliptic functions formalism. In terms of the latter
I derive in detail the classical background which solves the finite size
Euler-Lagrange equation and obtain the general path integral in finite size.
Both problem and solution offer an instructive
example of fruitful interaction between mathematics and physics.

\end{abstract}


\maketitle

\section*{I. Introduction}

The tunneling through a potential barrier plays a central role in
many areas of the physical sciences. It is well known that quantum
tunneling is an intrinsically nonlinear phenomenon which cannot be
attacked by standard perturbative methods. As an example, take a symmetric double
well potential (DWP) often encountered in the form of a two level system in condensed matter physics \cite{yuand,io4,io5}: the stable minima are
asymptotically connected by the classical (anti)kink solution, a
charge conserving domain wall, whose energy is inversely
proportional to the quartic force constant ($\delta > 0$) of the
$\phi^4$ potential at a given vibrational frequency inside the
well. Then, even small quartic nonlinearities induce large
classical energies which, in turn, enhance the effects of the
quantum fluctuations in the overall probability amplitude for a
particle to move from one minimum to the next. This explains why
standard perturbative techniques fail to describe tunneling
processes while semiclassical methods are known to provide an
adequate conceptual framework \cite{landau,berry}.

In general, $\phi^4$ models also host potentials having a negative
sign of the quadratic term ($\delta < 0$) \cite{sanchez}. In this
case the mid-point of the potential valley would be a classically
and locally stable configuration, a metastable state from which
the particle would tend to escape via quantum tunneling. In such
state, known in quantum field theory as "false vacuum"
\cite{coleman}, the system has a finite lifetime due to the
presence of a negative eigenvalue in the quantum fluctuation
spectrum which governs the decay into the "true vacuum". The
classical solution of the Euler-Lagrange equation for metastable
potentials is  a time-reversal invariant {\it bounce}
\cite{langer} carrying zero charge whose energy is inversely
proportional to the anharmonic parameter $\delta$. Then, also
analysis of metastable states and decay rates in tunneling
problems are unsuitable for perturbative methods.

Semiclassical methods usually extend the WKB
approximation \cite{garg} and permit to evaluate the
spectrum of quadratic quantum fluctuations around the classical
background. While such methods had been first applied to {\it infinite
size} systems, extensions to {\it finite size} systems have been later on
developed \cite{lusch} to obtain correlation functions and spectra
of $\phi^4$ and other model potentials \cite{vale03}. Classical
mechanical systems \cite{garcia,faris,burki} perturbed by
spatio-temporal noise are known to be sensitive to finite size
effects and, in particular, there is growing
interest in metallic nanowires and micromagnets
with magnetization reversal, in view of their technological
potential. Intense research is in progress also on biological systems and specifically DNA to establish the finite size effects on the physical properties \cite{joy2,manghi,i3}.

While a detailed study of decay rates in finite size {\it metastable
potentials} has been presented in Ref.\cite{i1}, this paper reconsiders the infinite size instantonic theory for a
{\it bistable $\phi^4$ potential}: in particular, I will point out the feature which justifies
a consistent finite size analysis
entirely based on the Jacobi elliptic functions
formalism. The latter is not exactly well known among the
physicists. A stimulating introduction to the elliptic functions
has been published by Good \cite{good} while a systematic
treatment can be found in the Refs. \cite{wang,whittaker}.

Thus, I feel, this is an interesting situation in which physics and mathematics mutually benefit by each other: the well known instanton problem offers the ground for an application of the Jacobian functions while the latter provides the fundamental tool to extend the instanton theory to cases of practical interest.

The key point of the instanton problem can be synthesized as follows:
due to the time translational
invariance a Goldstone mode appears in the system. As shown in
detail below, such mode is proportional to the (anti)kink velocity
and represents the ground state for the quantum fluctuation
spectrum. Accordingly, the associated zero mode eigenvalue
$\varepsilon_0$ breaks the Gaussian approximation and makes the
Euclidean path integral divergent. The trouble is overcome by
regularizing the fluctuations determinant and evaluating the
contribution of the zero mode separately. In doing that however,
the size $L$ of the system (the period of the instanton along the
time axis) is assumed large but {\it finite} so that
$\varepsilon_0^{-1/2} \propto L$ and the path integral is finite.
Then, to be consistent, also the classical equation of motion
should be solved assuming a finite $L$ and the (anti)kink,
peculiar of the infinite size theory, should be recovered as a
limit case of the more general, finite size (anti)instanton
solution. This problem is discussed in detail in Section II. In
Section III, I revisit the standard Euclidean path integral
\cite{feyn,fehi} implemented by the functional determinants method
which permits to derive, in the asymptotic limit, the contribution
stemming from the quantum fluctuations around the (anti)kink
\cite{kleinert,schulman}.  A consistent extension of the functional determinants theory to the finite size system allows one to obtain the path integral for the finite size instanton. The general path integral formula are given in Section IV while Section V contains
some final remarks.

\section*{II. Classical Equation of Motion}

Take a DWP described by:

\begin{eqnarray}
V(x)=\,- {{M\omega^2} \over 2}x^2 + {{\delta} \over 4}x^4 +
{{\delta} \over 4}a^4
\label{eq:1}
\end{eqnarray}

with minima located at $x=\, \pm a$ ($V(x=\,\pm a)=\,0$) and
positive $\delta$ in units $eV \AA^{-4}$. Note that the relation
$a=\, \sqrt{M\omega^2/ \delta}$ holds thus posing a constraint on
the potential parameters. Eq.~(\ref{eq:1}) is equivalent to:

\begin{eqnarray}
V(x)=\,{{M\omega^2} \over {4a^2}}\bigl(x - a \bigr)^2 \bigl(x + a
\bigr)^2 . \label{eq:2}
\end{eqnarray}

At a fixed $a$, a larger oscillator frequency $\omega$ inside the
well leads to an increased nonlinearity. Obviously there cannot be
classical motion through the barrier of a DWP and, in the real
time formalism, the very solutions are given by the vacua:
$x_{cl}=\,\pm a$. However, applying a Wick rotation which maps the
real time $t$ onto the imaginary time $\tau$, the DWP is turned
upside down and the classical equation of motion admits a non
trivial solution  \cite{jackiw}. The latter is the classical path
$x_{cl}(\tau)$ which minimizes the action in the imaginary time
path integral formalism. Let's find it solving the Euler-Lagrange
equation which reads:

\begin{eqnarray}
M\ddot{x}_{cl}(\tau)=\,V'(x_{cl}(\tau)) \label{eq:3}
\end{eqnarray}

where $V'$ means derivative with respect to $x_{cl}$. Integrating
Eq.~(\ref{eq:3}), one gets

\begin{eqnarray}
{M \over 2}\dot{x}_{cl}^2(\tau) - V(x_{cl}(\tau))=\, E
\label{eq:2a}
\end{eqnarray}

with integration constant $E$. Physically, we are looking for a
path covering the distance between $-a$ and $+a$ in the period
$L$. This means that the solution has to fulfill the boundary
conditions:

\begin{eqnarray}
{x}_{cl}(\tau=\,\pm L/2)=\, \pm a . \label{eq:2b}
\end{eqnarray}

Moreover, from Eq.~(\ref{eq:2a}) it is easily seen that the
relation

\begin{eqnarray}
{{\dot{x}_{cl}(\tau=\,\pm L/2)}}=\,\sqrt{2E/M} \label{eq:7a}
\end{eqnarray}

holds on the $\tau-$ range boundaries. Hence, only
(anti)instantons associated to values $E \geq 0$ can describe the
physics of the DWP. This marks an essential difference with
respect to the bounce solution of the Euler-Lagrange equation for
metastable potentials which, in fact, exists only for $E \leq 0$
\cite{i1}. Of course there is a deep
physical reason behind such difference in sign: the instantons
fulfills antiperiodic boundary conditions and carries finite
topological charge whereas the bounce is a time reversal invariant
object fulfilling periodic boundary conditions.

Integrating Eq.~(\ref{eq:2a}), one gets:

\begin{eqnarray}
\tau - \tau_0 =\,\pm \sqrt{{M \over 2}}
\int_{x_{cl}(\tau_0)}^{x_{cl}(\tau)} {{dx} \over {\sqrt{E +
V(x)}}} . \, \nonumber
\\ \label{eq:4}
\end{eqnarray}

$\tau_0$ represents the time at which the middle point of the
valley in the reversed potential is crossed. As the system is
translationally invariant, $\tau_0$ can be placed everywhere along
the imaginary time axis, $\tau_0 \in [-L/2, L/2]$. This
arbitrariness ultimately represents the physical origin of the
zero mode in the stability equation which governs the quantum
fluctuations spectrum (see below). Eq.~(\ref{eq:4}) admits a
solution which interpolates between the vacua providing evidence
for the existence of quantum tunneling. Such
solution is in fact a family of paths whose shape depends on the
associated energy $E$.

To be specific, it is clear from Eq.~(\ref{eq:4}) that the zero
energy configuration ($E=\,0$) is consistent with a physical
picture in which the particle starts at $x(\tau =\,- \infty)=\,
-a$ with zero velocity, falls in the valley and reaches the top of
the adjacent hill at $x(\tau =\,+ \infty)=\, +a$ with zero
velocity. Integration of Eq.~(\ref{eq:4}) with $E=\,0$ leads to
the classical path:

\begin{eqnarray}
x_{cl}(\tau)=\, \pm a \tanh[\omega (\tau - \tau_0)/\sqrt{2}] . \,
\nonumber
\\ \label{eq:4a}
\end{eqnarray}

This is the well known (anti)kink solution which exists if and
only if the potential minima are asymptotically connected. In the
spirit of the Matsubara formalism \cite{fehi,mahan}, the period
$L$ can be mapped onto the temperature axis as $L=\,\hbar/K_BT$.
Hence, for practical purposes, Eq.~(\ref{eq:4a}) holds in the {\it
low $T$ / large $L$} range whose {\it upper / lower} bound is set
by $\tau^*$ such that $|x_{cl}(\tau^*)|/a=\,1$. As long as the
latter condition is fulfilled the (anti)kink solution may be
assumed as a reliable approximation for the finite size
(anti)instanton. This terminology, born in quantum field theory
\cite{hooft,schaefer}, stems from the fact that the tunneling
transition happens in a short time, almost instantaneously. From
Eq.~(\ref{eq:4a}) one also notes that potentials with larger
$\omega$ sustain classical kinks in a broader temperature range.

Take now $E > 0$ and define:

\begin{eqnarray}
& &\chi_{cl}(\tau)=\, {1 \over {\sqrt{2}}}{{x_{cl}(\tau)} \over a}
\, \nonumber
\\
& &\kappa=\,{1 \over 2} + {{2E}\over {\delta a^4}} . \label{eq:5}
\end{eqnarray}

Then Eq.~(\ref{eq:4}) transforms into:

\begin{eqnarray}
\tau - \tau_0 =\,\pm {{1 \over \omega}}
\int_{\chi_{cl}(\tau_0)}^{\chi_{cl}(\tau)} {{d\chi} \over
{\sqrt{\chi^4 - \chi^2 + \kappa/2 }}} \label{eq:6}
\end{eqnarray}

which requires some work to be solved in terms of Jacobian elliptic
functions. The integral in the r.h.s. of Eq.~(\ref{eq:6}) yields:

\begin{eqnarray}
& &\int_{}^{} {{d\chi} \over {\sqrt{\chi^4 - \chi^2 + \kappa/2
}}}=\, {1\over 2} ^4\sqrt{2\over \kappa}F(\theta,s) \, \nonumber
\\
& &\theta=\,2\arctan\Bigl(^4 \sqrt{2\over \kappa} {}\chi_{cl}
\Bigr)\, \nonumber
\\
& &s^2=\,{1\over 2}\biggl(1 + {1 \over {\sqrt{ 2\kappa}}}\biggr).
\label{eq:22}
\end{eqnarray}

$F(\theta,s)$ is the elliptic integral of the first kind with
amplitude $\theta$ and modulus $s$:

\begin{eqnarray}
F(\theta,s)=\,\int_{0}^{\theta}{{d\alpha}\over {\sqrt{1 -
s^2\sin^2\alpha}}}, \label{eq:23}
\end{eqnarray}

whose quarter-period is given by the first complete elliptic
integral $K(s)=\, F(\pi/2,s)$ \cite{wang}. In terms of
$F(\theta,s)$, one defines the Jacobi elliptic functions
sn-amplitude, cn-amplitude, dn-amplitude :

\begin{eqnarray}
& & sn[F(\theta,s)]=\,\sin\theta\, \nonumber
\\
& & cn[F(\theta,s)]=\,\cos\theta\, \nonumber
\\
& & dn[F(\theta,s)]=\,\sqrt{1 - s^2 \sin^2\theta}. \label{eq:24}
\end{eqnarray}

From Eqs.~(\ref{eq:6}),~(\ref{eq:22}) one gets:

\begin{eqnarray}
\pm  ^4\sqrt{\kappa \over 2 }2{{\omega}}(\tau - \tau_0)
=\,F(\theta,s) \label{eq:25}
\end{eqnarray}

and using the relations:

\begin{eqnarray}
& &\tan\theta=\,{{2\tan(\theta/2)} \over {1 - \tan^2(\theta/2)}}
\, \nonumber
\\
& &\tan\theta= {{sn[F(\theta,s)]} \over {cn[F(\theta,s)]}},
\label{eq:26}
\end{eqnarray}

Eq.~(\ref{eq:25}) transforms into:

\begin{eqnarray}
& &{{sn(2\varpi,s)} \over {cn(2\varpi,s)}}= \,{{2 \cdot \,{}
^4\sqrt{2\over \kappa} {}\chi_{cl}} \over {1 - \sqrt{2\over
\kappa} {}\chi_{cl}^2}}\, \nonumber
\\
& &\varpi=\,\pm ^4\sqrt{\kappa \over 2 }{{\omega}}(\tau - \tau_0).
\label{eq:27}
\end{eqnarray}

The condition $^4 \sqrt{2\over \kappa} {}\chi_{cl}(\tau) \neq \pm
1$ imposed by Eq.~(\ref{eq:27}) is equivalent to $x_{cl}(\tau)/a
\neq \pm (2s^2-1)^{-1/2}$ which is always fulfilled in the
physical range of our interest. The l.h.s. in the first of
Eq.~(\ref{eq:27}) can be rewritten using the double arguments
relations:

\begin{eqnarray}
& &sn(2\varpi,s)=\,{{2sn(\varpi,s)cn(\varpi,s)dn(\varpi,s)} \over
{1 - s^2sn^4(\varpi,s)}} \, \nonumber
\\
& &cn(2\varpi,s)=\,{{cn^2(\varpi,s) -
sn^2(\varpi,s)dn^2(\varpi,s)} \over {1 - s^2sn^4(\varpi,s)}} \,
\nonumber
\\
\label{eq:28}
\end{eqnarray}

and, through the definition in Eq.~(\ref{eq:5}), from
Eqs.~(\ref{eq:27}),~(\ref{eq:28}) I derive the two solutions for
the classical equation of motion in the finite size model:

\begin{eqnarray}
& &x^{(1)}_{cl}(\tau)=\, ^4\sqrt{2\kappa}\,{} a
{{sn(\varpi,s)dn(\varpi,s)} \over {cn(\varpi,s)}}\, \nonumber
\\
& &x^{(2)}_{cl}(\tau)=\,- ^4\sqrt{2\kappa}\,{} a {{cn(\varpi,s)}
\over {sn(\varpi,s)dn(\varpi,s)}}. \, \nonumber
\\
\label{eq:8}
\end{eqnarray}

Set now $E=\, 0$ in Eq.~(\ref{eq:8}). For $\kappa=\,1/2$,
$s^2=\,1$ and the elliptic functions reduce to the limit cases
\cite{abram}:

\begin{eqnarray}
& & sn(\varpi,1)=\,tanh \varpi \, \nonumber
\\
& & cn(\varpi,1)=\,sech \varpi \, \nonumber
\\
& & dn(\varpi,1)=\, sech \varpi . \label{eq:28a}
\end{eqnarray}

Then, from Eq.~(\ref{eq:8}), I get consistently:

\begin{eqnarray}
& &x^{(1)}_{cl}(\tau) \rightarrow \, \pm a \tanh[\omega (\tau -
\tau_0)/\sqrt{2}] \, \nonumber
\\
& &x^{(2)}_{cl}(\tau) \rightarrow \, \mp a \coth[\omega (\tau -
 \tau_0)/\sqrt{2}] . \label{eq:10}
\end{eqnarray}

The first in Eq.~(\ref{eq:10}) is the (anti)kink of
Eq.~(\ref{eq:4a}) whose center of motion is set at
$x^{(1)}_{cl}(\tau_0)=\,0$. Instead, the second in
Eq.~(\ref{eq:10}) is the solution of Eq.~(\ref{eq:4}) fulfilling
the condition $x^{(2)}_{cl}(\tau_0)=\,\infty$. This proves the
correctness of the analytical work.

Computation of
Eq.~(\ref{eq:8}) is an instructive task. Numerical investigation shows that both backgrounds
predict the same physics. This is checked by establishing the one
to one correspondence between $E$ and the temperature $T^*$ at
which the boundary conditions expressed by Eq.~(\ref{eq:2b}) are
fulfilled: setting a value for $E$, both classical backgrounds
fulfill Eq.~(\ref{eq:2b}) at the same $T^*$.

Note that both solutions present singularities. The zeros of
$x^{(1)}_{cl}(\tau)$, occurring at $\varpi=\, \pm 2nK(s)$ with
integer $n$  would be divergencies for $x^{(2)}_{cl}(\tau)$.
Viceversa, the zeros of $x^{(2)}_{cl}(\tau)$ at $\varpi=\, \pm (2n
+ 1)K(s)$ are singular points for $x^{(1)}_{cl}(\tau)$.

For practical purposes, the choice of the first in
Eq.~(\ref{eq:8}) turns out to be more convenient. In fact, taking
the center of motion at $x_{cl}(\tau_0=\,0)=\,0$,
$x^{(1)}_{cl}(\tau)$ properly describes a family of finite size
(anti)instantons which are bounded, continuous and odd in the
range $\tau \in [-L/2,L/2]$.

The boundary conditions (see Eq.~(\ref{eq:2b})) for $x^{(1)}_{cl}(\tau)$
are always fulfilled {\it before} encountering the singular points,
thus the paths are always continuous functions within their
oscillation periods \cite{i1}. Mathematically this means that,
defining $\varpi^*=\,\pm ^4\sqrt{\kappa \over 2 }{{\omega}}L/2$,
one finds numerically $\varpi^*/ K(s) < 1$, hence the {\it time
period} $L$ is covered within a {\it lattice period} shorter than
$2K(s)$.

Around $x^{(1)}_{cl}(\tau)$, one can further study the finite size
effects on the quantum fluctuation spectrum by means of the theory
of the functional determinants. The details are given in
Refs.\cite{i2} and \cite{kirsten1} while, in the next Section, I review  the general
features of the functional determinants method and evaluate the
quantum fluctuations contribution for the case of the infinite
size theory.

\section*{III. Quantum Fluctuations around the (anti)kink}

For a particle with mass $M$ moving in a potential $V(x(\tau))$
the probability amplitude to propagate, in the imaginary time $L$,
from the initial position $x_i$ (at $-L/2$) to the final position
$x_f$ (at $L/2$) is given by the path integral \cite{feyn}

\begin{eqnarray}
<x_f |x_i>_L=\,\int Dx \exp\Bigl[- {{A[x]} \over {\hbar}}
\Bigr]\, \nonumber
\\
A[x]=\,\int_{-L/2}^{L/2} d\tau \biggl({M \over 2} \dot{x}(\tau)^2
+ V(x(\tau)) \biggr) , \, \nonumber
\\ \label{eq:30}
\end{eqnarray}

where $A[x]$ is the Euclidean action that weighs the contribution
of the generic path $x(\tau)$. $Dx$ is the measure of the path integration \cite{gelfand}. Accordingly, by
expanding the Euclidean action in the neighborhood of the
classical path, the propagation amplitude for a particle in the
bistable quartic potential reads

\begin{eqnarray}
& &<x_f |x_i>_L=\,\exp\biggl[- {{A[x_{cl}]} \over {\hbar}} \biggr]
\int D\eta \exp\biggl[- {{A_f[\eta]} \over {\hbar}} \biggr]
\, \nonumber
\\
& &A[x_{cl}]=\,\int_{-L/2}^{L/2} d\tau \biggl({M \over 2}
\dot{x}_{cl}^2(\tau) +  V(x_{cl}(\tau))\biggr) \, \nonumber
\\
& &A_f[\eta]=\,\int_{-L/2}^{L/2} d\tau \biggl({M \over 2}
\dot{\eta}^2(\tau) + {1 \over 2}V''(x_{cl}(\tau))\eta^2(\tau)
\biggr) \, \nonumber
\\
& &V(x_{cl}(\tau))=\,{{M\omega^2} \over {4a^2}}\bigl(x_{cl}(\tau)
- a \bigr)^2 \bigl(x_{cl}(\tau) + a \bigr)^2  \, \nonumber
\\
& &V''(x_{cl}(\tau))=\,{{M\omega^2} \over
{a^2}}\bigl(3x_{cl}^2(\tau) - a^2 \bigr). \, \nonumber
\\ \label{eq:31}
\end{eqnarray}

$D\eta$, defined below, is the integration measure over the path fluctuations.
If $L \rightarrow \infty$, from Eq.~(\ref{eq:4a}) and the second
of Eq.~(\ref{eq:31}), one obtains the classical path action:

\begin{eqnarray}
A[x_{cl}]&=&\, {{2 \sqrt{2}} \over 3} {{M^2 \omega^3} \over
\delta}. \label{eq:32}
\end{eqnarray}

This is also the (anti)kink energy in units of $\hbar$ exhibiting
the peculiar dependence on $\delta$ emphasized in the
Introduction.

Taking the $\delta \rightarrow 0$ limit (for a given $\omega$)
corresponds to decouple the two potential wells ($a \rightarrow
\infty$). Accordingly the classical energy required to connect the
minima becomes infinitely large. A slightly different version of
$V(x_{cl}(\tau))$ in Eq.~(\ref{eq:31}) is often found in the
literature with $\omega$ replaced by $\omega/\sqrt{2}$. In this
case the action would be given by Eq.~(\ref{eq:32}) without the
$\sqrt{2}$ factor.

The quadratic fluctuations contribution to the Euclidean action is
derived by the third and fifth of Eq.~(\ref{eq:31}). Performing an
integration by parts on the kinetic term, the quantum fluctuation
action transforms into:

\begin{eqnarray}
& &A_f[\eta]=\,{M \over 2} \int_{-L/2}^{L/2} d\tau \Bigl(
-{{d^2{\eta}(\tau)}\over {d\tau^2}} +\, \nonumber
\\
& &+ 2\omega^2 \bigl(1 - {3 \over 2} \cosh^{-2}\varpi
\bigr)\eta(\tau) \Bigr) \eta(\tau) \, \nonumber
\\
& &\varpi=\,\omega (\tau - \tau_0)/\sqrt{2} . \label{eq:33}
\end{eqnarray}

The infinite size background of Eq.~(\ref{eq:4a}) has been
inserted in Eq.~(\ref{eq:33}). Then, consistently, Dirichlet
boundary conditions $\eta(\tau=\,\pm L/2)= 0$, with $L \rightarrow
\,\infty$, have to be assumed to derive Eq.~(\ref{eq:33}) from
Eq.~(\ref{eq:31}). I'll come back to this point later. Expanding
$\eta(\tau)$ in a series of ortonormal components $\eta_n(\tau)$
with coefficients $\varsigma_n$,

\begin{eqnarray}
& &\eta(\tau)=\,\sum_{n=\,0}^{\infty} \varsigma_n \eta_n(\tau)\,
\label{eq:34}
\end{eqnarray}

one observes that the quantum fluctuation components obey a
Schr\"{o}dinger equation with Rosen-Morse potential

\begin{eqnarray}
& &\hat{O} \eta_n(\tau)= \, \varepsilon_n \eta_n(\tau)\, \nonumber
\\ & &\hat{O}=\,-{{d^2}\over {d\tau^2}} + 2\omega^2 \bigl(1 - {3 \over 2}
\cosh^{-2}\varpi \bigr) \, \nonumber
\\
\label{eq:35}
\end{eqnarray}

which admits two bound state eigenvalues ($\varepsilon_0,\,
\varepsilon_1$) and a continuum spectrum ($\varepsilon_n,\, n\geq
2$) \cite{landau}. In particular, the ground state $\eta_0(\tau)$
is the Goldstone mode, discussed in the Introduction, whose
eigenvalue is $\varepsilon_0=\,0$.

Thus, after solving Eq.~(\ref{eq:35}), the quantum action formally
reads:

\begin{eqnarray}
A_f[\eta]=\,{M \over 2} \sum_{n=\,0}^{\infty} \varepsilon_n
\varsigma^2_n .  \label{eq:36}
\end{eqnarray}

Note that, dimensionally: $[\varepsilon_n]\equiv [sec^{-2}]$,
$[\eta_n]\equiv [sec^{-1/2}]$, $[\varsigma_n]\equiv [\AA
\,sec^{1/2}]$.

Using the measure \cite{schulman}:

\begin{eqnarray}
\int D\eta=\,\aleph \prod_{n=\,0}^{\infty}
\int_{-\infty}^{\infty} {{d\varsigma_n}\over {\sqrt{2\pi\hbar/M}}}
\, \nonumber
\\ \label{eq:37}
\end{eqnarray}

and performing Gaussian integrations over the {\it directions}
$\varsigma_n$ one formally obtains the quantum fluctuations
contribution to the propagator in Eq.~(\ref{eq:31}):

\begin{eqnarray}
\int D\eta \exp\biggl[- {{A_f[\eta]} \over {\hbar}}
\biggr]=\,{\aleph \Bigl( \prod_{n=\,0}^{\infty}\varepsilon_n
\Bigr)^{-1/2}}. \, \nonumber
\\ \label{eq:38}
\end{eqnarray}

The normalization constant $\aleph$ accounts for the Jacobian in
the transformation to the normal mode expansion in
Eq.~(\ref{eq:34}).

There is a singularity in Eq.~(\ref{eq:38}) caused by the
$\varepsilon_0=\,0$ eigenvalue which breaks the Gaussian
approximation. This reflects the arbitrariness in the choice of
$\tau_0$ in Eq.~(\ref{eq:4a}). The latter represents in fact a one
parameter family of solutions localized around their time center
and, as such, they lack the time translational symmetry of the
Hamiltonian. Any solution, in the set given by Eq.~(\ref{eq:4a}),
can undergo an infinitesimal time translation which corresponds to
an infinitesimal shift of $\tau_0$. The resulting shifted kink is
however degenerate with the original one and therefore such a fluctuation
costs no action. This is the physical origin of the zero mode.
Accordingly we have the recipe to handle the divergence in
Eq.~(\ref{eq:38}): it can be easily seen that $d x_{cl}(\tau)/d
\tau$ satisfies the Euler-Lagrange equation and solves the
homogeneous differential problem associated to Eq.~(\ref{eq:35}),
$\hat{O}\eta_n(\tau)=\,0$. Moreover it has no nodes being
$x_{cl}(\tau)$ monotonic. Then, $d x_{cl}(\tau)/d \tau$ is
proportional to the Goldstone mode $\eta_0(\tau)$ of the
ortonormal expansion in Eq.~(\ref{eq:34}).

These physical facts permit to tackle the divergence. In fact the
integration measure in Eq.~(\ref{eq:37}) has a contribution along
the $\varsigma_0$ direction. Thus, transforming from the
$\varsigma_0$ to the $\tau_0$-axis (with the norm of $d
x_{cl}(\tau)/d \tau$ as Jacobian) one gets \cite{schulman}

\begin{eqnarray}
& & \int_{-\infty}^{\infty} {{d\varsigma_0}\over
{\sqrt{2\pi\hbar/M}}}\exp\biggl({{-M\varepsilon_0 \varsigma_0^2}
\over {2\hbar}} \biggr)=\,\sqrt{{{M }\over
{2\pi\hbar}}}\bar{N}^{-1} \int_{-L/2}^{L/2} {d\tau_0}\, \nonumber
\\
& & (\varepsilon_0)^{-1/2} \rightarrow \sqrt{{{M}\over
{2\pi\hbar}}}\bar{N}^{-1}L\, \nonumber
\\
& & \bar{N}^{-1}=\,\sqrt{ 2\int_0^{L/2}d\tau
|\dot{x}_{cl}(\tau)|^2}.
\label{eq:39}
\end{eqnarray}

The replacement $(\varepsilon_0)^{-1/2} \rightarrow \,
\sqrt{{{A[x_{cl}]}/ {2\pi\hbar}}}L$ is often encountered in the
literature. The latter however holds only in the $L\rightarrow \infty$ limit for which
$A[x_{cl}]\equiv \,M \bar{N}^{-2}$ whereas, for finite size/temperature systems, the replacement in
Eq.~(\ref{eq:39}) is correct in general.

Anyway the heart of the matter, unambigously shown by
Eq.~(\ref{eq:39}), is that the divergence in the path integral can
be removed only by keeping $L$ {\it finite}. It can be certainly large but it has to remain {\it finite}. On the other hand the (anti)kink
solution in Eq.~(\ref{eq:4a}) is intrinsically based on the
assumption $L \rightarrow \infty$. Moreover, as emphasized above,
its derivative ($\propto \eta_0(\tau)$) satisfies Dirichlet
boundary conditions only at $L \rightarrow \infty$. One may face
the contradiction by forcing the fluctuation to vanish in a
finite $\tau-$ range: this would compress the particle
distribution thus lifting the $\varepsilon_0$ eigenvalue upwards
and removing the zero \cite{langer}. Although the discrepancy
(with respect to the asymptotic picture) is of order
$O(\exp(-\omega L))$ and therefore it may be controlled at large
$L$, I felt that there was a need for a consistent formulation of
the theory able to incorporate finite size effects already at the
classical stage. This justifies the work presented in the previous
Section.

Hereafter I complete the derivation of the path integral based on
the infinite size (anti)kink. Thus, Eq.~(\ref{eq:38}) can be
rewritten as:

\begin{eqnarray}
& &\int D\eta(\tau) \exp\biggl[- {{A_f[\eta]} \over {\hbar}}
\biggr]=\,{{F(\omega L)}\over {(\varepsilon_0
\varepsilon_1)^{1/2}}} {{\Bigl(
\prod_{n=\,0}^{\infty}\varepsilon^\omega_n \Bigr)^{1/2}} \over
{{\Bigl( \prod_{n=\,2}^{\infty}\varepsilon_n
\Bigr)_{cont}^{1/2}}}} \, \nonumber
\\
& &F(\omega L)=\,{{\aleph}\over
{\Bigl({\prod_{n=\,0}^{\infty}\varepsilon^\omega_n\Bigr)^{1/2}}}}. \,
\nonumber
\\ \label{eq:40}
\end{eqnarray}

Formally, I have multiplied and divided by the product of the
harmonic quantum fluctuation eigenvalues $\varepsilon^\omega_n$
\cite{kleinert}. $F(\omega L)$ is the well known harmonic
fluctuation factor which absorbs $\aleph$ \cite{kleinert}. The
two bound states have been factored out so that the set of
continuum eigenvalues (denominator) is discrete but it contains
two states less than the harmonic spectrum (numerator). Then, the
evaluation of the quadratic quantum fluctuation contribution is
practically reduced to the estimate of the {\it ratio} of
functional determinants in Eq.~(\ref{eq:40}). This has been done
by several methods in the literature
\cite{mckane,jafari,kleinert2} extending fundamental ideas by
Gelfand and Yaglom \cite{gelfand} and Forman \cite{forman}. The
fact that a {\it ratio} of determinants appears is not a simple
mathematical trick. In fact each determinant, being the product
over an infinite number of eigenvalues with magnitude greater than
one, would diverge in the $T \rightarrow 0$ limit. Such
(exponential) divergences cancel out in the ratio which is
meaningful both in value and sign and arises naturally in the path
integral method \cite{gelfand}.

In the instantonic approach, the essential observation is that
both determinants satisfy the Jacobi equations (that is, the
homogeneous differential equations associated to the respective
stability equations) around their classical solutions, see
Ref.\cite{schulman} for an extensive discussion. The remarkable
path integral result  which follows is that the ratio is
determined only on the base of the classical action (or of the
classical path velocity) and it yields a factor $9\omega^4$. Then,
including also the bound state eigenvalue $\varepsilon_1=\,3
\omega^2/4$ in Eq.~(\ref{eq:40}) and taking the square root of the
whole, the fluctuations contribution gets:

\begin{eqnarray}
\int D\eta(\tau) \exp\biggl[- {1 \over {\hbar}} A_f[\eta]
\biggr]=\,2 \sqrt{3} \omega F(\omega L) \sqrt{{{M
\bar{N}^{-2}}\over {2\pi\hbar}}}L . \, \nonumber
\\ \label{eq:41}
\end{eqnarray}

This result completes the calculation of the quantum fluctuation for the infinite size case
in the semiclassical approximation of Eq.~(\ref{eq:31}).

\section*{IV. Infinite and Finite Size Path Integrals}

Assume now that, in the propagation amplitude, the initial and
final positions are very close to the potential minima: $x_i \sim
-a$, $x_f \sim a$ and say $\varphi_0(x \pm a)$ the ground state
harmonic wavefunctions in the potential wells. For $x \simeq \pm
a$, $\varphi_0(x + a)\varphi_0(x - a) \sim \sqrt{{M\omega}\over
{\pi \hbar}}$. The harmonic factor, $F(\omega L)$ can be
approximated in the large $L$ limit as, $F(\omega L)\sim
\,\sqrt{{M\omega}\over {\pi \hbar}}\exp(-\omega L/2)$. Then, from
Eq.~(\ref{eq:31}) and Eq.~(\ref{eq:41}), the path propagator for a
particle in a single double well potential reads:

\begin{eqnarray}
& &<x_f | x_i >_{L \rightarrow \infty}=\,\sqrt{{M\omega}\over {\pi \hbar}} \exp{(-\omega L/2)} \Omega_\infty L \, \nonumber
\\
& &\Omega_\infty=\, \exp\biggl[- {{A[x_{cl}]} \over {\hbar}}  \biggr] 2
\sqrt{3} \omega \sqrt{{{M \bar{N}^{-2}}\over {2\pi\hbar}}}. \,
\nonumber
\\ \label{eq:42}
\end{eqnarray}

This is the general result of the infinite size instantonic
theory. The quantum tunneling removes the two fold ground state
degeneracy and induces a splitting in the energy levels. The size
of this quantum effect is governed by the classical action hosted
in the expression of $\Omega_\infty$. The classical path solution
determines the shape of the potential in the stability equation
(Eq.~(\ref{eq:35})) thus providing the frame within which the
quadratic quantum fluctuations display their strength and set the
tunneling rate.

Finally and fundamentally, note that $\Omega_\infty$ does not depend on
temperature in the infinite size instantonic approach as
$A[x_{cl}]$ is a constant in such limit. This limitation of the
theory is removed in the finite size method based on the classical
background of Section II.

Around such background one has to evaluate the quantum fluctuations contribution by finding the functional determinants {\it ratio} for a finite size system \cite{mckane} after extracting the zero mode as shown above.
Working out the calculations, I get the semiclassical path integral for one
(anti)instanton in finite size $L$:

\begin{eqnarray}
& &<x_f | x_i>_L=\,\sqrt{{M \over {2\pi \hbar L}}} \cdot {1 \over
{2 \sinh\bigl(\omega L/{2}\bigr)}} \Omega(L) L \, \nonumber
\\
& &\Omega(L)=\,\exp\biggl[- {{A[x_{cl}]} \over {\hbar}}  \biggr]
\cdot \Omega^{QF}(L)  \, \nonumber
\\
& &\Omega^{QF}(L)=\, \sqrt{{{M \bar{N}^{-2}}\over {2\pi\hbar}}}
\sqrt{\Bigl|{{Det[\hat{h}]} \over {Det^R [\hat{O}]}}\Bigr|} \label{eq:21}
\end{eqnarray}

with:

\begin{eqnarray}
& &Det[\hat{h}]=\,-4\sinh^2(\omega L/2) \, \nonumber
\\
& &\hat{h}\equiv \, -\partial^2_{\tau} + \omega^2
\, \nonumber
\\
& &
Det^R [\hat{O}]=\, \prod_{n=\,1}^{\infty}\varepsilon_n  . \label{eq:41}
\end{eqnarray}

Note that the harmonic determinant has been, in turn,
normalized over the free particle determinant
$Det[-\partial^2_{\tau}]=\, \sqrt{{M \over {2 \pi \hbar L}}} $
which incorporates the constant $\aleph$.
In the $L \rightarrow \infty$ limit, I get

\begin{eqnarray}
{{Det[\hat{h}]} \over {Det^R[\hat{O}]}} \rightarrow {{12
\omega^2}} \label{eq:43a}
\end{eqnarray}

thus recovering the value of the infinite size
instantonic approach given in Eq.~(\ref{eq:42}). This proves the correctness of the
analytical procedure.

$\Omega^{QF}(L)$ in Eq.~(\ref{eq:21}) accounts for the quantum fluctuation effects in
the finite size tunneling problem while $\Omega(L)$ is the
overall tunneling frequency which removes the twofold degeneracy
of the double well potential in a finite domain. From the path
integral in Eq.~(\ref{eq:21}) one can extract the physical
properties of the closed quantum system for any size.

\section*{V. Final Remarks}

I have reviewed the instantonic method for the propagation
amplitude of a particle moving via tunneling in a bistable quartic
potential. This has been done within the standard semiclassical
scheme for a $\phi^4$ model admitting the (anti)kink solution as
classical path which interpolates between the vacua in a {\it
infinite time}. Since the time is conceived as imaginary by a Wick
rotation, an {\it infinite time} description maps onto the zero
temperature limit according to the Matsubara formalism. The
classical action and the stability equation for the quantum
fluctuations have also been solved in the asymptotic limit.
However, in deriving the particle propagator I have pointed out
that, at some steps, the time (the {\it size} of the system) is
in fact assumed to be {\it large but finite} in order to tackle the
divergency due to the quantum fluctuation zero eigenvalue related
to the translational invariance of the system. Led by this
observation, I have described the general method which permits a
consistent extension of the instantonic approach such to
incorporate finite size effects both in the classical background
and in the quantum spectrum. The fundamental step, the replacement
of the {\it infinite size} (anti)kink by the {\it finite size}
classical path of Eq.~(\ref{eq:8}), has been explained in
detail making use of the formalism of the Jacobi elliptic
functions. Along similar patterns one generalizes the bounce
solution of the Euler-Lagrange equation for metastable potentials
to the {\it finite size} case \cite{i1}. As a general consequence
of the finite size formalism, the
quantum fluctuation spectrum obtained in semiclassical
approximation becomes dependent on the system size or, in the
Matsubara formalism, on the system temperature. From a physical
point of view, the elliptic functions formalism permits to
evaluate the size/temperature effects both on the tunneling energy
of bistable potentials and on the decay rate in systems with
metastable potentials.

\end{document}